\documentclass[
prd,
superscriptaddress,
10 pt,
preprintnumbers,
notitlepage,
secnumarabic,
nobibnotes,
nofootinbib,
showpacs
]{revtex4-1}

\usepackage[T1]{fontenc}
\usepackage{bm}
\usepackage[pdftex]{color,graphicx}
\usepackage[makeroom]{cancel}
\usepackage{amssymb}
\usepackage{mathtools}
\usepackage{tabularx}
\usepackage{dsfont}
\usepackage[artemisia]{textgreek}
\usepackage{nicefrac}
\usepackage{hyperref}
\usepackage{xcolor}
\usepackage{tikz}
\usetikzlibrary{shapes}
\usetikzlibrary{arrows}
\tikzset{l/.style={draw=black, line width=1pt}}
\tikzset{s/.style={auto, outer sep=-1}}
\tikzset{t/.style={auto, outer sep=2, pos=-0.1}}
\tikzset{b/.style={auto, outer sep=2, pos=1.1}}
\tikzset{z/.style={auto, outer sep=-2}}
\tikzset{y/.style={auto, outer sep=1}}
\tikzset{v/.style={auto, outer sep=-2}}
\tikzset{h/.style={auto, outer sep=1}}
\tikzset{a/.style={pos=1.1}}
\tikzset{l1/.style={draw=black, line width=0.1pt}}
\tikzset{l2/.style={draw=black, line width=0.2pt}}
\tikzset{l4/.style={draw=black, line width=0.4pt}}
\tikzset{l6/.style={draw=black, line width=0.6pt}}

\usepackage{perpage}
\MakePerPage{footnote}
\newcommand{\kbar}{\mathchar'26\mkern-9mu k}

\newcommand{\ket}[1]{\ensuremath{\left|#1\right\rangle}}
\newcommand{\bracket}[2]{\ensuremath{\left\langle #1 \middle| #2 \right\rangle}}

\DeclareMathAlphabet\mathbfcal{OMS}{cmsy}{b}{n}

\begin{document}
\title{Quantum Reduced Loop Gravity with matter:\\eigenvectors of the Hamiltonian operator in isotropic cosmology}

\author{Jakub Bilski}
\email{jakubbilski14@fudan.edu.cn}
\affiliation{Center for Field Theory and Particle Physics \& Department of
Physics, Fudan University, 200433 Shanghai, China}

\author{Antonino Marcian\`o}
\email{marciano@fudan.edu.cn}
\affiliation{Center for Field Theory and Particle Physics \& Department of
Physics, Fudan University, 200433 Shanghai, China}


\begin{abstract}
\noindent  Introducing a new method, we demonstrate how the action of reduced operators can be derived without resorting to a recoupling theory and how they exactly reproduce the results obtained in the standard approach of Quantum Reduced Loop Gravity (QRLG). This is particularly relevant while dealing with volume operator when dealing with the coupling of matter fields to gravity. Apart from reinforcing the close link between QRLG and loop quantum cosmology (LQC), this procedure also sheds new light on the issue of how to extract the continuum limit, without resorting to the large-$j$ expansion, thereby pointing towards a new approach to tackle this problem.
\end{abstract}
\maketitle

	
\section{Introduction}\label{I}
\noindent
Within the framework of Loop Quantum Gravity (LQG) \cite{Ashtekar:2004eh,Rovelli:2004tv,Thiemann:2007zz}, several strategies have been deployed hitherto in order to capture the cosmological sector of the theory. Notably Loop Quantum Cosmology (LQC) was developed to deal with the quantization of the symmetry-reduced phase space of the full theory, namely LQG, taken into account --- see \textit{e.g.} \cite{Bojowald:2011zzb,Ashtekar:2011ni,Banerjee:2011qu}. Other approaches naturally followed, with the purpose of linking LQC to the full theory of LQG, since quantization and symmetry-reduction need not, a priori, commute. Several possibilities, which were investigated, exists within the literature. The possibility that quantization and reduction commute was studied in \cite{Brunnemann:2007du, Engle:2013qq, Hanusch:2013jza, Hanusch:2013jza, Fleischhack:2015nda}, unravelling how the quantum configuration spaces of LQC can be embedded into the full theory. Light on the use of spinfoam techniques was shed in \cite{Rovelli:2009tp}, while coherent state techniques were proposed within the Group Field Theory approach in \cite{Gielen:2013naa,Oriti:2016qtz}.

Quantum reduced loop gravity (QRLG) is chronologically one of the latest attempts, having been proposed in \cite{Alesci:2012md} and then developed in \cite{Alesci:2013xd, Alesci:2013xya, Alesci:2014uha} (for a review see \cite{Alesci:2016gub}). (Soon after QRLG was proposed, a `gauge unfixing' procedure, which is closely related to the implementation of Dirac Brackets and the quantization of the classically gauge fixed reduced phase space, was suggested in \cite{Bodendorfer:2014vea}.) It relies on imposing weak gauge-fixing conditions to the states of the kinematical Hilbert space of the full theory, LQG. This peculiarity allows one to recover the cosmological sector directly from LQG. Classically, the gravitational systems considered are those ones described by metrics with spatial part and dreibein gauge-fixed to a diagonal form. Gauge-fixing conditions are then applied weakly on the kinematical Hilbert space of the full theory. As a result, Bianchi I models can be successfully recovered in this framework. Furthermore, within the semiclassical limit, QRLG reproduces the effective Hamiltonian of LQC \cite{Alesci:2014uha,Alesci:2014rra}, in the $\mu_0$ regularization scheme. It is also worth mentioning that the effective improved dynamics \cite{Ashtekar:2006wn} can be inferred by averaging over the ensemble of the classically equivalent states \cite{Alesci:2016rmn}.

For these reasons QRLG is thought to provide a novel derivation of earlier results of LQC, including the realization of the singularity-resolution scenario. Nonetheless, the quantum gravity scenario can be completed only once matter fields are taken into account. Since QRLG introduces a graph structure underlying the description of the continuous universe at the classical level, and since the origin of the discretization must be recovered at the quantum level, the quantization of the matter fields will be achieved via the same tools of LQG \cite{Thiemann:1997rq,Thiemann:1997rt}. As a consequence, QRLG offers a framework to test implications of loop quantization for matter fields. The first analysis focusing on scalar matter field was developed in \cite{Bilski:2015dra}. The implementation of gauge vector fields was then developed in \cite{Bilski:2016pib,Bilski:2017dgi}, keeping in mind the potential role of vector fields in cosmology \cite{Ford:1989me, Parker:1993ya, Golovnev:2008cf, Maleknejad:2011sq, Alexander:2011hz, Adshead:2012kp, Alexander:2014uza}. This has particular relevance for quantum gravity phenomenology, and the possibility of developing unified formalisms for all forces, with peculiar phenomenological consequences \cite{Alexander:2011jf, Alexander:2012ge}.

In this work we also come back to the link between LQC and QRLG, developing an alternative perspective about their relation. We indeed define the basic canonical variables along the edges and vertices of the fundamental cuboids we adopt in the tessellation, and directly apply methods borrowed from LQC to derive the regularized Hamiltonian operator within the framework of RLQG. The implementations of these new techniques and the development of new tools that are closely related to LQC has also the advantage of simplifying the coupling of matter fields to gravity. For instance, it is much more straightforward to deal with the volume operator in this framework than in the original formulation of RLQG. Furthermore, we get further insight into tackling how to recover the continuum limit in this formalism.

The paper is organized as follows. In Sec.~\ref{II}, we minimally couple the standard model of particle physics to gravity, cast the theory within the Hamiltonian formalism in terms of real (Ashtekar-Barbero) gravitational variables, and recover the kinematical Hilbert space of the theory. As a by-product of quantum gauge fixing, in Sec.~\ref{III}, we present the reduced lattice representation of the fundamental fields, including reduction of the Ashtekar variables. In Sec.~\ref{IV} we show how the actions of reduced operators can be derived without a recoupling theory and how they reproduce exactly the results obtained in the standard approach of RLQG. In Sec.~\ref{V} we discuss results obtained and argue on the relevance of the new method for QRLG.


Through the paper we will use the convention on the metric signature $(-,+,+,+)$, we will introduce the reduced gravitational coupling constant $\kbar=\frac{1}{2}\gamma\hbar\kappa=8\pi\gamma l_P^2$ ($\gamma$ and $l_P$ are the Immirzi parameter and the Planck length respectively), and set the speed of light to $c=1$. The metric tensor is defined as $g_{\mu\nu}=e^I_{\mu}e^J_{\nu}\eta_{IJ}$, where $e^I_{\mu}$ are vierbein fields and $\eta_{IJ}$ being the flat Minkowski metric. The spatial metric tensor reads $q_{ab}=e^i_a e^j_b\delta_{ij}$, where $e^i_a$ denotes dreibeins. The lowercase latin indexes $a,b,...=1,2,3$ label coordinate on each Cauchy hypersurface constructed by ADM decomposition \cite{Arnowitt:1962hi}, while $i,j,..=1,2,3$ are $\mathfrak{su}(2)$ internal
indexes and $\delta_{ij}$ stands for the Kronecker delta. Notice that $\mathfrak{su}(2)$
generators are defined as $\tau_i=-\frac{i}{2}\sigma_i$, where $\sigma_i$ are Pauli matrices. Lie algebra indexes for a generic gauge group G are labeled by $A,B,...=1,2,...,$ dim(G). Finally, indexes $\alpha,\beta,...$ represent internal fermionic transformation in some irreducible representation of the gauge group under consideration. Indexes written in the bracket $(\,)$ are not summed, while for every other repeated pair the Einstein convention is applied.


\section{Free fundamental fields in the formalism of LQG}\label{II}

\noindent
We start by minimally coupling the standard model of particle physics to gravity, and cast the theory within the Hamiltonian ADM formalism in terms of real, \textit{i.e.} Ashtekar-Barbero, gravitational variables. We then recover the kinematical Hilbert space of the theory, including matter fields degrees of freedom.

\subsection{Classical Gravity: Fixing our notations}\label{II.1}

\noindent
We define the Einstein-Hilbert action with the cosmological constant term, minimally coupled to the free fields of the Standard Model,
\begin{equation}\label{action_with_matter}
S:=S^{(g)}+S^{(\Lambda)}+\!\int_M\!\!\!\!d^{4}x\sqrt{-g}\,\mathcal{L}^{(\text{matter})},
\end{equation}
where $\mathcal{L}^{(\text{matter})}$ encodes the Yang-Mills field, the scalar field, and the Dirac field. 

In this section, we focus on the first two terms of $S$. The starting point in the construction of the Hamiltonian operator in LQG would be the Einstein-Hilbert action, which reproduces the classical equation of motion,
\begin{equation}\label{action_EH}
S^{(g)}+S^{(\Lambda)}:=\frac{1}{\kappa}
\int_M\!\!\!\!d^{4}x\sqrt{-g}R-\frac{\Lambda}{\kappa}\int_M\!\!\!\!d^{4}x\sqrt{-g},
\end{equation}
where $R$ is the Ricci scalar. For completeness, we have also added the cosmological constant $\Lambda$ in the action. Here $g$ stands for the determinant of the metric tensor $g_{\mu\nu}$ and the gravitational coupling constant reads $\kappa=16\pi G$.

The canonical quantization procedure in LQG is realized on the Hamiltonian obtained from action $S^{(g)}$, which is derived in the ADM formalism \cite{Arnowitt:1962hi}
while using the Ashtekar variables \cite{Ashtekar:1986yd}. The latter are the Ashtekar-Barbero connection $A^i_a=\Gamma^i_a+\gamma K^i_a$ and the densitized dreibein $E_i^a=\sqrt{q}e_i^a$. Here, $\Gamma^i_a := \frac{1}{2}\epsilon^{ijk}\Gamma_{jka}= -\frac{1}{2}\epsilon^{ijk}e_k^b(\partial_ae^j_b-\Gamma^c_{ab}e^j_c)$
is the spin connection and $\gamma K^i_a=\Gamma^i_{\ 0a}$ is the extrinsic
curvature. They form a canonically conjugate pair of variables, with a Poisson structure given by
\begin{equation}
\left\{A^i_a(t,{\bf x}), E_j^b(t,{\bf
y})\right\}=\gamma\frac{\kappa}{2}\delta_a^b\,\delta_j^i\,\delta^{(3)}({\bf
x}-{\bf y})\,.
\end{equation}
An important remark is that the Ashtekar variables are introduced by a canonical point transformation on the gravitational phase space from the ADM canonical variables when the latter are written in the first-order form $(K^i_a,E_i^a)$. From now on, for consistency of notation, we will use superscript ${}^{(A)}$ rather than ${}^{(g)}$ in order to denote objects describing gravitational degrees of freedom. Since we foliate spacetime and restrict our analysis to three-dimensional spatial hypersurfaces with metric tensor $q_{ab}$ on it, we reserve the term `metric' only to this object.

The Hamiltonian, which is obtained by the Legendre transform of \eqref{action_EH}, reads
\begin{equation}
H^{(A)}_T+H^{(\Lambda)}_T
=\int_{\Sigma_t}\!\!\!\!d^3x\,\Big(
A^i_t\mathcal{G}_i^{(A)}
+N^a\mathcal{V}_a^{(A)}+N\big(\mathcal{H}^{(A)}+\mathcal{H}^{(\Lambda)}\big)\Big),
\end{equation}
where the three elements
\begin{align}
\label{A_Gauss}
G^{(A)}:=&\ \frac{1}{\gamma\kappa}\!\int_{\Sigma_t}\!\!\!\!d^3x\,A^i_t
D_aE^a_i,
\\
\label{A_vector}
V^{(A)}
:=&\ \frac{1}{\gamma\kappa}\!\int_{\Sigma_t}\!\!\!\!d^3x\,N^a
F^i_{ab}E^b_i,
\\
\phantom{\hspace{-49mm}}\text{and}\phantom{\hspace{49mm}}\nonumber
\\
\label{A_scalar}
H^{(A)}+H^{(\Lambda)}
:=&\ \frac{1}{\kappa}\!\int_{\Sigma_t}\!\!\!\!d^3x\,N
\bigg(
\frac{1}{\sqrt{q}}\big(F^i_{ab}-(\gamma^2+1)\epsilon_{ilm}K^l_aK^m_b\big)\epsilon^{ijk}E_j^aE_k^b
+\Lambda\sqrt{q}
\bigg)
\end{align}
are called respectively the Gauss, the diffeomorphism (or vector) and the
Hamiltonian (or scalar) constraints. They impose respectively an internal SU$(2)$, spatial diffeomorphism and a time reparametrization invariance. Hence the Hamiltonian constraint describes dynamics on the SU$(2)$ and the spatial diffeomorphism (or in short, diffeomorphism) invariant subspace. Objects $(A^i_t,N^a,N)$ are Lagrange multipliers. The quantity $F^i_{ab}$ denotes the curvature of the Ashtekar connection, while $D_a$ is a metric and dreibein compatible covariant derivative.

Lattice regularization in LQG is performed in two steps. In the first step,
we begin from the imposition of the so-called `Thiemann trick', which goes as
\begin{align}
\label{E_trick}
\frac{1}{E_i^a}\big(\sqrt{|E|}\big)^{\!n}
=&\ \frac{2}{n}\frac{\delta\mathbf{V}^{n}}{\delta E_i^a}
=\frac{4}{n\gamma\kappa}\big\{A^i_a,\mathbf{V}^{n}\big\}\,,
\\
K^i_a=&\ \frac{\delta\mathbf{K}}{\delta
E^a_i}=\frac{2}{\gamma\kappa}\big\{A^i_a,\mathbf{K}\big\},
\end{align}
where $|E|=q$ is the absolute value of the determinant of $E^a_i$ and
$\mathbf{K}=\int\!d^3xK^i_aE^a_i$. This latter object, a densitized integral of the extrinsic curvature, is precisely known as the `Thiemann complexifier' in the literature.

The second step is to granulate the spatial hypersurfaces. It is realized by a
construction of small solid objects --- grains, which fill all of the spacelike Cauchy hypersurface and intersect each other only in lower-dimensional submanifolds. This
granulation of space is regulated by the parameter $\varepsilon$. The limit $\varepsilon\to0$ corresponds to the granulating object of trivial volume or, in other words, corresponds to taking the regulator to zero. This is done in a way similar to taking the decoupling limit in effective field theories --- decreasing the volume of the grains while at the same time increasing their the number in the way such that they always fill out the full space. The standard choice for the shape of the solids is a tetrahedron and then the procedure is called a triangulation. The detailed description of this method can be found in \cite{Thiemann:1996aw,Thiemann:2007zz}. An alternative, much simpler choice for the shape of the solids is a cube --- albeit resulting in fixing some of the gauge freedom of the theory. This is the case of the `cubulation' procedure used in QRLG \cite{Alesci:2013xd}.

As a consequence of the granulation of the space, we get a natural regularization of the dynamical variables. The remarkable result worth mentioning is that after quantization, the effect of the regularization is to remove both gravitational singularities (the initial singularity in a classical cosmology and the black hole singularity) and the UV-singularities of quantum matter fields \cite{Thiemann:2007zz}. Finally, the identification between the space and a graph $\Gamma$ that is created as a consequence of the granulation is realized by a duality: $\Gamma$ consists of links and nodes, hence in the dual graph we get respectively faces and volumes of the grains of $\Gamma^*$.\\

At the level of the canonical variables, the regularization is realized as follows. The Ashtekar connection $A^i_a$ is recovered from holonomies, or parallel transports of the connection, along links $l^a$,
\begin{equation}
h_{l}\!:=\mathcal{P}\exp\!\left(i\int_{l}A^j_a(l(s))\tau^j\dot{l}^a(s)\right).
\end{equation}
Consequently, the curvature of the connection $F^i_{ab}$ is turned into the holonomy around loop $a\circlearrowleft b$ that starts from the initial point of link $l^a$, goes along this link and through the shortest polygon chain, it returns along link
$l^b$ to the initial point. This is realized via the following relations:
\begin{align}
\label{connection_to_holonomy}
h_{l^p}^{-1}\big\{h_{l^p},{\bf V}\big\}=&\
\varepsilon\big\{A_a,\mathbf{V}\big\}\delta^a_p+\mathcal{O}(\varepsilon^2),
\\
\label{curvature_to_holonomy}
2\epsilon^{pqr}h_{q\circlearrowleft r}=&\ \epsilon^{pqr}\big(h_{q\circlearrowleft
r}-h_{q\circlearrowleft r}^{-1}\big)
=\epsilon^{ab(c)}\varepsilon^2F_{ab}\,\delta^p_c+\mathcal{O}(\varepsilon^4),
\end{align}
where $F_{ab} = F^j_{ab}\tau_j$, $A_a = A_a^i\tau_i$ and $p,q,r...$ label
directions of the links of $\Gamma$.
As a result we obtain the scalar constraint density written in terms of $h_{l^p}$, $h_{q\circlearrowleft r}$, $\mathbf{V}$ and (in the case of the gravitational term) $\mathbf{K}$, namely
\begin{equation}
\begin{split}\label{g_lattice_scalar}
H^{(A)}
=&\
\frac{1}{\kappa}\lim_{\varepsilon\to0}\frac{1}{\varepsilon^3}\!\int\!\!d^3x\,N\epsilon^{pqr}\bigg(
\frac{2^3}{\gamma\kappa}\,\text{tr}\Big(h_{p\circlearrowleft
q}\,h_{l^r}^{-1}\big\{\mathbf{V},h_{l^r}\big\}\Big)
\\
&-\frac{2^5(\gamma^2+1)}{\gamma^3\kappa^3}
\,\text{tr}\Big(
h_{l^p}^{-1}\big\{\mathbf{K},h_{l^p}\big\}\,h_{l^q}^{-1}\big\{\mathbf{K},h_{l^q}\big\}\,h_{l^r}^{-1}\big\{\mathbf{V},h_{l^r}\big\}
\Big)
\bigg).
\end{split}
\end{equation}
Having the lattice-regulated scalar constraint, the quantization method is straightforward and can be implemented \textit{via} the Dirac procedure. We turn Poisson brackets into commutators multiplied by $1/i\hbar$ and change the dynamical variables into operators.

The kinematic Hilbert space of LQG (or its reduced equivalents, like QRLG) is the direct sum of cylindrical functions of (possibly reduced, \textit{i.e.} diagonal) connections along the links of the graph $\Gamma$ (the general graph or cuboidal one ${}^R\Gamma$ in the reduced case). In the case of LQG, the kinematic Hilbert space is equipped with an inner product defined as an integral over cylindrical functions with a SU$(2)$-invariant Haar measure. In the Alesci-Cianfrani model the SU$(2)$ group is restricted to three U$(1)$ subgroups defined along the directions of basis vectors that diagonalize the $\mathfrak{su}(2)$ generators $\tau^i$.

The basis states, called spin network states $\bracket{h}{\Gamma;j_l,i_v}$, are
labeled with a graph $\Gamma$, along with spin $j_l$ of the holonomy attached to each link $l$ of $\Gamma$ through irreducible representations of $\mathfrak{su}(2)$ --- these are Wigner matrices in the case of standard LQG --- and with intertwiner $i_v$ implementing SU$(2)$ invariance at each node $v$ of $\Gamma$. In QRLG the basis states are ${}_{\raisebox{-1pt}{\scriptsize$R$\!\!}} \left<h\middle|\{\Gamma,j_l,i_v\}\right>_{\!R}$ and involve Wigner matrices, which are rotated to the directions that diagonalize
$\tau^i$ and then projected on coherent Livine-Speziale states \cite{Livine:2007vk}, with maximal or minimal spin number $j_l$ --- consequently, the fundamental representation takes values $\pm1/2$.

The kinematical Hilbert space is constructed as a space of solutions of the constraints \eqref{A_Gauss} and \eqref{A_vector}. In the case of the Alesci-Cianfrani model, the SU$(2)$ invariance is replaced with the three U$(1)$ symmetries along the directions of the links of ${}^R\Gamma$, while the diffeomorphism constraint is restricted to the implementation of an invariance under spatial diffeomorphisms, which do not generate any off-diagonal components. A precise definition of the Hilbert space in full LQG can be found \textit{\textit{e.g.}} in \cite{Thiemann:2007zz}, while that for QRLG is given \textit{\textit{e.g.}} in
\cite{Alesci:2013xd}. 

Finally, the problem of solving the Hamiltonian constraint at the quantum level recasts
as the problem of finding solutions of the action $\hat{H}\ket{\Gamma;j_l,i_v}$ --- in the reduced case, the action $^{R\!}\hat{H}\ket{\Gamma;j_l,i_v}_{\!R}$.\\

While taking into account the cosmological constant's sector, the whole difficulty in finding a solution to the equation $H^{(\Lambda)}\ket{\Gamma;j_l,i_v}$ becomes the derivation of the action of the volume operator,
\begin{equation}\label{volume}
\hat{\mathbf{V}}
\ket{\Gamma;j_l,i_v}.
\end{equation}
The gravitational Hamiltonian $H^{(A)}$ produces two classes of equations for the eigenvalues of the $\mathfrak{su}(2)$ traces of the operators in \eqref{g_lattice_scalar}.  As usual in the standard literature, we are going to call the first one the Euclidean term. It reads
\begin{equation}\label{loop}
\,\text{tr}\Big(\hat{h}_{p\circlearrowleft
q}\,\hat{h}_r^{-1}\,\hat{\mathbf{V}}_{\!v}\,\hat{h}_r\Big)
\ket{\Gamma;j_l,i_v}\,,
\end{equation}
where $\hat{h}_r:=\hat{h}_{l^r}$, with $l^r$ being the lattice link along direction $r$. The second, being the most complicated object, has been named the Lorentzian term and it is given by the formula
\begin{equation}\label{curvature}
\,\text{tr}\Big(\hat{h}_p^{-1}\,\hat{\text{K}}_v\,\hat{h}_p\,\hat{h}_q^{-1}\,\hat{\text{K}}_v\,\hat{h}_q
\,\hat{h}_r^{-1}\,\hat{\mathbf{V}}_{\!v}\,\hat{h}_r\Big)
\ket{\Gamma;j_l,i_v}.
\end{equation}

It is worth noting that equation \eqref{volume} is solvable for simple
configurations of states. However a problem that arises is the fact that there is
an ambiguity in the choice of the definition of the volume operator \cite{Lewandowski:1996gk,Flori:2008nw}. Besides that, as we will
see in the next sections, in order to derive actions of the complete set of all
the Standard Model matter fields, we need rather some powers of
$\hat{\mathbf{V}}_{\!v}$. Hence, instead of focusing only on the cosmological
constant sector described by formula \eqref{volume}, we need to solve the
following action,
\begin{equation}\label{volume_n}
\big(\hat{\mathbf{V}}_{\!v}\big)^{\!n}
\ket{\Gamma;j_l,i_v},
\end{equation}
$n$ being a positive real number.

In the case of equation \eqref{loop}, the solutions for standard LQG has been
found only for single-node states of a particular valency\footnote{
See \textit{e.g.} \cite{Borissov:1997ji,Gaul:2000ba,Alesci:2013kpa} (for the trivalent nodes) or \cite{Alesci:2011ia} (for the tetravalent nodes).}.
However in the case of the reduced
graph, a general solution exists, as shall be elaborated later in section \eqref{III}.

Derivation of the Lorentzian term equation \eqref{curvature} is even more
demanding. As in the case of the Euclidean term, the result for a general case
with a big number of nodes of different valency is rather impossible to be
achieved. However, in the reduced model this term does not appear any more. This
is a consequence of the diagonalization of the spatial metric tensor.

\subsection{Standard Model of particles}\label{II.2}

\noindent
The generally covariant action for free fields that are components of the Standard Model of particle physics can be written as
\begin{equation}\label{SM_action}
\begin{split}
S^{(\text{matter})}=&\ S^{(\underline{A})}+S^{(\phi)}+S^{(\Psi)}=
\!\int_M\!\!\!\!d^{4}x\sqrt{-g}\bigg(
-\frac{1}{4Q^2}
\Big(
g^{\mu\nu}g^{\xi\pi}\underline{F}^A_{\mu\xi}\underline{F}^A_{\nu\pi}-2m^2g^{\mu\nu}\underline{A}_{\mu}^A\underline{A}_{\nu}^A
\Big)
\\&
+\frac{1}{2\zeta}\Big(
g^{\mu\nu}\big(\underline{\nabla}_{\mu}\phi^A\big)\big(\underline{\nabla}_{\nu}\phi^A\big)-\mu^2\phi^A\phi^A
\Big)
+\frac{i}{2}\Big(
\overline{\Psi}\gamma^Ie_I^{\mu}\underline{\nabla}_{\mu}\Psi
-\overline{\underline{\nabla}_{\mu}\Psi}\gamma^Ie_I^{\mu}\Psi
\Big)\bigg).
\end{split}
\end{equation}
Here $Q^2$ and $\zeta$ are coupling constants, $\mu$ is  an inverse of the Compton wavelength of the Klein-Gordon field, while $m$ is  an inverse of the Compton wavelength of the Proca field (in the Yang-Mills case, we simply put $m=0$). $\underline{F}^A_{\mu\xi}$ is a curvature of an SU$(N)$ gauge potential $\underline{A}_{\mu}^A$, $\underline{\nabla}_{\mu}$ is an $\text{SL}(2,\mathds{C})\times\text{G}$ covariant derivative --- acting on the gravitational degrees of freedom as a $\text{SL}(2,\mathds{C})$ connection --- that annihilates the metric tensor $g_{\mu\nu}$, vierbein $e_I^{\mu}$ and gamma matrix $\gamma^I$, while $G$ denotes the gauge group. Notice that $\underline{\nabla}_{\mu}$ acts as
\begin{equation}
\underline{\nabla}_{\mu}\phi_A=\underline{D}_{\mu}\phi_A=\partial_{\mu}\phi_A+\underline{f}_{ABC}\underline{A}_{B\mu}\phi_C
\end{equation}
and
\begin{equation}
\underline{\nabla}_{\mu}\Psi=\underline{D}_{\mu}\Psi+\frac{1}{4}\text{\it\textGamma}_{\mu
IJ}\gamma^{[I}\gamma^{J]}\Psi =\partial_{\mu}\Psi+\underline{\tau}_AA_{\mu}^A\Psi+\frac{1}{4}\text{\it\textGamma}_{\mu IJ}\gamma^{[I}\gamma^{J]}\Psi\,,
\end{equation}
where $\underline{D}_{\mu}$ denotes an SU$(N)$ covariant derivative, while $\text{\it\textGamma}_{\mu IJ}$ is a Lorentz spin connection including torsional degrees of freedom. The SU$(N)$-valued scalar field has been denoted by $\phi^A$, while $\Psi=(\psi,\eta)$ is the Dirac bispinor.  It is worth mentioning that cubic and quartic interactions are constructed only form the terms already included in \eqref{SM_action}.

The Legendre transform of the action $S=S^{(g)}+S^{(\Lambda)}+S^{(\text{matter})}$ --- we included the gravitational terms for completeness --- reads
\begin{equation}\label{T_Hamiltonian}
H_T=H^{(A)}_T+H^{(\Lambda)}_T+H^{(\underline{A})}_T+H^{(\phi)}_T+H^{(\Psi)}_T
=\int_{\Sigma_t}\!\!\!\!d^3x\,\Big(
A^i_t\mathcal{G}_i^{(A)}+\underline{A}^A_t\underline{\mathcal{G}}_A
+N^a\mathcal{V}_a+N\mathcal{H}\Big),
\end{equation}
where the constraints have the following forms:
\begin{align}
\label{T_g_Gauss}
G=&\ G^{(A)}+G^{(\Psi)}=\!\int_{\Sigma_t}\!\!\!\!d^3x\,A^j_t
\bigg(
\frac{1}{\gamma\kappa}\mathcal{D}_aE^a_j+i\big(\xi^{\dagger}\tau_j\xi-\chi^{\dagger}\tau_j\chi\big)
\bigg),
\\
\label{T_Gauss}
\underline{G}=&\ \underline{G}^{(\underline{A})}+\underline{G}^{(\phi)}
=-\!\int_{\Sigma_t}\!\!\!\!d^3x\,\underline{A}^A_t
\big(
\underline{D}_a\underline{E}^a_A+\underline{f}_{ABC}\,\phi^B\pi^C
\big),
\\
\label{T_vector}
V=&\ V^{(A)}+V^{(\underline{A})}+V^{(\phi)}+V^{(\Psi)}
=\!\int_{\Sigma_t}\!\!\!\!d^3x\,N^a
\bigg(
\frac{1}{\gamma\kappa}F^i_{ab}E^b_i+\underline{F}^A_{ab}\underline{E}^b_A+\pi^A\underline{D}_a\phi^A
+\frac{i}{2}\big(\xi^{\dagger}\mathcal{D}_a\xi-\chi^{\dagger}\mathcal{D}_a\chi-\text{c.c.}\big)
\bigg),
\\
\begin{split}
\label{T_scalar}
H=&\ H^{(A)}+H^{(\Lambda)}+H^{(\underline{A})}+H^{(\phi)}+H^{(\Psi)}
\\
=&\int_{\Sigma_t}\!\!\!\!d^3x\,N
\bigg(
\frac{1}{\kappa}\frac{1}{\sqrt{q}}\big(F^i_{ab}-(\gamma^2+1)\epsilon_{ilm}K^l_aK^m_b\big)\epsilon^{ijk}E_j^aE_k^b
+\frac{\Lambda}{\kappa}\sqrt{q}
\\
&+\frac{Q^2\!}{2\sqrt{q}}q_{ab}
\big(
\underline{E}^a_A\underline{E}^b_A+\underline{B}^a_A\underline{B}^b_A
\big)
+\frac{\sqrt{q}}{2Q^2}q^{ab}m^2\underline{A}_{a}^A\underline{A}_{b}^A
\\
&+\frac{\zeta}{2\sqrt{q}}\pi^A\pi^A
+\frac{\sqrt{q}}{2\zeta}q^{ab}\underline{D}_a\phi^A\,\underline{D}_b\phi^A
+\frac{\sqrt{q}}{2\zeta}\mu^2\phi^A\phi^A
\\
&+\frac{E_j^a}{2\sqrt{q}}\Big(
\mathcal{D}_a\big(\xi^{\dagger}\sigma_j\xi+\chi^{\dagger}\sigma_j\chi\big)
+i\big(\xi^{\dagger}\sigma_j\mathcal{D}_a\xi-\chi^{\dagger}\sigma_j\mathcal{D}_a\chi-\text{c.c.}\big)
-K^j_a\big(\xi^{\dagger}\xi-\chi^{\dagger}\chi\big)
\Big)
\bigg).
\end{split}
\end{align}
Notice that the $\underline{G}$, which does not appear before, is the Gauss constraint imposing the SU$(N)$ invariance. Spinors $\psi,\eta$ have been replaced with the half-pseudoscalars --- the scalar densities of weight one half --- namely $\xi=\sqrt[4]{q}\,\psi$ and $\chi=\sqrt[4]{q}\,\eta$. This redefinition is allowed since covariant derivative $\mathcal{D}_a\psi=(\partial_a+\tau_iA^i_a)\psi$ annihilates $q$ and the Legendre transform is unaffected by this change due to the relation
$\sqrt{q}(\psi^{\dagger}\dot{\psi}-\dot{\psi}^{\dagger}\psi)=\xi^{\dagger}\dot{\xi}-\dot{\xi}^{\dagger}\xi$.

The total Hilbert space of the matter-gravity interactions reads
\begin{equation}
\mathcal{H}_{kin}^{(tot)}=\mathcal{H}_{kin}^{(A)}\!\otimes\mathcal{H}_{kin}^{(\text{matter})},
\end{equation}
where $\mathcal{H}_{kin}^{(\text{matter})}$ is the tensor product of the Hilbert spaces of the representations of different matter fields --- for detailed definitions of these Hilbert spaces see \cite{Thiemann:1997rt,Thiemann:1997rq,Bilski:2015dra,Bilski:2016pib,Bilski_Thesis}. The basis states are represented by
$\ket{\Gamma;j_l,i_v;\text{matter}}=\ket{\Gamma;j_l,i_v}\otimes\ket{\Gamma;\text{matter}}$,
where $\ket{\Gamma;\text{matter}}$ is the tensor product of the states in $\mathcal{H}_{kin}^{(\text{matter})}$, containing only matter degrees of freedom.

The dynamics of the system is encoded in the Hamiltonian constraint $H$. With the procedure of canonical quantization this object is replaced by the operator acting on the basis state,
\begin{equation}
\hat{H}\ket{\Gamma;j_l,i_v;\text{matter}}.
\end{equation}
All the matter fields are turned into operators. The gravitational degrees of freedom coupled with matter are encoded in the following objects:
\begin{align}
q_{ab}/\sqrt{q}=&\ \sqrt{|E|}E_a^iE_b^i,
\\
1/\sqrt{q}=&\ \sqrt{|E|}E_a^iE_b^ie^a_je^b_j/3,
\\
\sqrt{q}\,q^{ab}=&\
\sqrt{|E|}E^i_{a'}E^{i'}_{b'}\frac{1}{4}\epsilon_{ijk}\epsilon^{acd}e^j_ce^k_d\epsilon_{i'lm}\epsilon^{bef}e^l_ee^m_f,
\\
\sqrt{q}=&\ \sqrt{|E|}E_a^iE_b^i
\frac{1}{12}\epsilon_{i'jk}\epsilon^{acd}e^{i'}_{a'}e^j_ce^k_d\epsilon_{j'lm}\epsilon^{bef}e^{j'}_{b'}e^l_ee^m_f
\\
E_i^a/\sqrt{q}=&\ \frac{1}{2}\epsilon_{ijk}\epsilon^{abc}\sqrt{|E|}E_b^jE_c^k,
\end{align}
where $E^i_a=(E_i^a)^{-1}$.
Notice that the identities above allow to define diffeomorphism invariant representations for each single matter field in $\hat{H}$ --- see \textit{e.g.} \cite{Thiemann:2007zz,Bilski_Thesis}. If we will release this condition, we can simplify the metric tensor identities, still keeping the diffeomorphism invariant restriction on the full Hamiltonian. Anyway, we obtain combinations of quantity $E^i_a\big(\sqrt{|E|}\big)^{\!n}$ that can be replaced with $4\big\{A^i_a,\mathbf{V}^{n}\big\}/(n\gamma\kappa)$ using the Thiemann's trick \eqref{E_trick}. It is worth mentioning that for the diffeomorphism invariant representations for each matter field (of weight density 1), $n$ always equals $1/2$. However, in the case of different representations (see \textit{e.g.} \cite{Bilski:2015dra}), or the cubic or quartic interactions, $n$ takes different values \cite{Bilski_Thesis}.

As a result, all the gravitational degrees of freedom coupled with lattice representations of the matter fields, lead to the operator action
\begin{equation}\label{coupling}
\,\text{tr}\Big(\tau^i\hat{h}_p^{-1}\hat{\mathbf{V}}_{\!v}^n\,\hat{h}_p\Big)
\ket{\Gamma;j_l,i_v}.
\end{equation}

\section{Reduced lattice representation of fundamental fields}\label{III}

\noindent
As was already mentioned in the previous section, formulas (\ref{loop}), (\ref{curvature}), (\ref{volume_n}) and (\ref{coupling}), which are un-tractable in the general case of LQG, have analytical solutions in QRLG.

Let us begin with the volume operator. The regularized action of this operator $\eqref{volume_n}$ in full LQG reads
\begin{equation}\label{volume_operator}
\begin{split}
\big(\hat{\mathbf{V}}_{\!v}\big)^{\!n}
\ket{\Gamma;j_l,i_v}
&=
\bigg(\frac{1}{3!}\!\int\!\!d^3x
\Big(\Big|
\epsilon^{ijk}\epsilon_{pqr}\hat{E}_i(S^p)\hat{E}_j(S^q)\hat{E}_k(S^r)
\Big|\Big)^{\!\!\frac{1}{2}}
\bigg)^{\!\!n}
\ket{\Gamma;j_l,i_v},
\end{split}
\end{equation}
where it has been assumed that the operator of a volume to a given power equals that power of the volume operator. The irregularity and complication of a structure of the general graph directly prevents from getting solution to equation \eqref{volume_operator}. Since the same operator appears in other equations such as \eqref{loop}, \eqref{curvature} and \eqref{coupling}, it follows that these cannot be solved either. The situation is much simpler in the Alesci-Cianfrani model with the regular cuboidal, self-dual\footnote{The self-duality of $\Gamma$ should be understood in a geometrical way. The faces dual to the links of $\Gamma$ and the polyhedra dual to the nodes of this graph are respectively squares and cubes. They are elements of the dual graph $\Gamma^*$, which is congruent to $\Gamma$.} graph. \\

Classical models involving a diagonal spatial metric describe a wide class of physical objects, including Kerr-Newman black holes and Bianchi I Universe, where the restriction on the symmetry does not break spatial diffeomorphism invariance. In these physically relevant models, quantum implementation of the diagonal gauge has to be done simultaneously at the level of the Hilbert space, and of all the constraints of the full generally covariant LQG. This provides an independence of the background metric as well as the gauge invariance of the operators that carry dynamical degrees of freedom and could be interpreted as observables. These properties are the quantum equivalents of the general covariance, which classically is encoded in the equivalence principle.

As a by-product quantum gauge fixing, \textit{i.e.} restricting the spatial part of the metric to be diagonal, results in picking a cuboidal structure of graph $\Gamma\to{}^{R}\Gamma$ and in reducing SU$(2)$ symmetry of states in U$(1)^3$ symmetry --- each U$(1)$ along one internal direction. The diagonal form of the spatial metric can be realized at the level of the Ashtekar variables by the definitions
\begin{align}
\label{r_A}
{}^R\!A^i_a(t,{\bf x}):=&\ \frac{1}{l_0}c_{(i)}(t,{\bf x})\,\delta^i_a,
\\
\label{r_E}
{}^R\!E_i^a(t,{\bf x}):=&\ \frac{1}{l_0^2}p^{(i)}(t,{\bf x})\,\delta^a_i,
\end{align}
where $l_0$ is the length of the side of the fiducial cuboidal graphs we introduced. Notice that these reduced phase space variables satisfy the canonical relation in the fiducial volume $l_0^3$ of the basic grain of space,
\begin{equation}
\big\{c_i(t,{\bf x}),p^j(t,{\bf y})\big\}=\frac{\kappa\gamma}{2}\delta_i^j\delta^{(3)\!}({\bf x}-{\bf y}).
\end{equation}

In QRLG, the volume operator \eqref{volume_operator} has been first derived in the gravitational sector in the Alesci-Cianfrani model \cite{Alesci:2013xd}. The eigenvalue of the reduced operator of the volume of a region ${\bf R}^3$ reads
\begin{equation}\label{red_volume}
\begin{split}
{}^{R}\hat{\mathbf{V}}^n\big({\bf R}^3\big)
\ket{\Gamma;j_l,i_v}_{\!R}
=&
\sum_{v\in{\bf
R}^3}\!\big(\big|\hat{p}^1(v)\hat{p}^2(v)\hat{p}^3(v)\big|\big)^{\!\frac{n}{2}}\!
\ket{\Gamma;j_l,i_v}_{\!R}
=
\!\sum_{v\in{\bf R}^3}\!\hat{\mathbf{V}}^n
\ket{c_v;j_l,i_v}_{\!R}
\\
=&
\sum_{v\in{\bf
R}^3}\!\big(\big|\kbar^3\,\Sigma_v^{(1)}\,\Sigma_v^{(2)}\,\Sigma_v^{(3)}\big|\big)^{\!\frac{n}{2}}\!
\ket{\Gamma;j_l,i_v}_{\!R}
=:
\!\sum_{v\in{\bf R}^3}\!\big({}^{R}\mathbf{V}_{\!v}\big)^{\!n}
\ket{\Gamma;j_l,i_v}_{\!R}\,,
\end{split}
\end{equation}
where $\Sigma^{(i)}_{v}:=\frac{1}{2}\big(j^{(i)}_{v}+j^{(i)}_{v-\vec{e}_i}\big)$
denotes the mean value of the spin along a direction $i$. Notice that the volume
operator has a discrete spectrum, with the minimal eigenvalue in the case of the
fundamental representation of the spin corresponding to $\Sigma^{(i)}_{v}=1/2$. The fiducial length $l_0$ then has a cut-off at the quantum level on the limit of the lattice parameter $\varepsilon$, which can be identified roughly with $\varepsilon_0 = \sqrt{\kbar/2}$. Assuming that space is discretized with a minimal length, corresponding to the cube root of the minimal value of the eigenvalue of ${}^{R}\hat{\mathbf{V}}$, then the following relation holds $\varepsilon_0=l_0$. This is a crucial conceptual point in our analysis: although the cuboids create a fiducial granular structure only for the purposes of regularization of the theory, here we identify the regulator to a physical scale which gives rise to a fundamental nonlocality on quantum scales. In other words, we identify a physical non-zero parameter to the minimum scale to which the regulator can be shrunk, setting which to zero shall reproduce the continuum limit\footnote{Finding the continuum limit of the theory is obviously a more subtle issue and shall be addressed in a future work \cite{Bilski_et_al.}}.

In \eqref{red_volume} we introduced a new object, $\ket{c_v;j_l,i_v}_{\!R}$, which is called a basic cell of the cuboidal graph ${}^{R}\Gamma$. The term $8c_v$ individuates a region around the hexavalent node $v$, which is surrounded by three perpendicular pairs of collinear links --- one incoming and one outgoing --- that emanate from $v$ and end or begin in nearest neighbor nodes. Node $v$ is at the center of $c_v$, which is a cube normalized by a factor $1/8$, having faces with centers positioned on the nearest neighbor nodes. The cellular decomposition of the eigenstates of the volume operator can be naturally extended to the eigenstates of other operators appearing in \eqref{volume}, \eqref{loop} and \eqref{coupling}. Since this triple and the volume operator form the complete set of dynamical operators in QLRG, the cellular decomposition can be implemented also on the reduced kinematical Hilbert space ${}^{R}\mathcal{H}_{kin}^{(tot)}$, yielding
\begin{equation}\label{sum_QRLG}
{}^{\text{R}\!}\hat{O}\ket{\Gamma;j_l,i_v}_{\!R}=\sum_{v\in\Gamma}{}^{\text{R}\!}\hat{O}_v\ket{\Gamma;j_l,i_v}_{\!R},
\end{equation}
where ${}^{R}\hat{O}:=\Big(
{}^{R}\hat{\mathbf{V}},
\,\text{tr}\big({}^{R}\hat{h}_{i\circlearrowleft j}\,{}^{R}\hat{h}_k^{-1}\,
{}^{R}\hat{\mathbf{V}}\,{}^{R}\hat{h}_k\big),
\,\text{tr}\big(\tau^i\,{}^{R}\hat{h}_j^{-1}\,{}^{R}\hat{\mathbf{V}}^n\,{}^{R}\hat{h}_j\big)
\Big)$. Note that the Lorentzian term described by \eqref{curvature} is missing in this expression. This is due to the fact that using the diagonal, reduced variables \eqref{r_A} and\eqref{r_E}, this term equals zero. As a result of the cellular structure of ${}^{R}\mathcal{H}_{kin}^{(tot)}$, we can restrict calculations to the single-cell action,
\begin{equation}\label{cell_QRLG}
{}^{\text{R}\!}\hat{O}\ket{c_v;j_l,i_v}_{\!R}=\hat{O}_v\ket{\Gamma;j_l,i_v}_{\!R},
\end{equation}
which, for simplicity, is what we shall assume from now on. Finally, note that the self-duality of ${}^{R}\Gamma$ introduces the same cellular structure to the space $\big({}^{R}\mathcal{H}_{kin}^{(tot)}\big)^{\!*}$.

\section{New method of calculation HCO for reduced models}\label{IV}

\subsection{Gravitational sector}\label{IV.1}
\noindent
In this section we show how the actions of reduced operators ${}^{R}\hat{O}$ can
be derived without a recoupling theory and how they reproduce exactly the results
obtained in the standard approach \cite{Alesci:2015nja,Bilski:2015dra,Bilski:2016pib}.\\

The lattice representations of reduced gravitational variables $c_{(i)}\delta^i_p$ and $p^{(i)}\delta^i_p$ acting along link $l^p$ are the following. The reduced connection is replaced with a U$(1)$ holonomy defined as
\begin{equation}
{}^{R}h_{l^p}=e^{\pm\varepsilon {}^{\,R\!}A_{(k)\!}\tau^k}\delta^k_p\,.
\end{equation}
The reduced momentum is smeared through surface $\mathbf{S}^p$ normal to link $l^p$ of fiducial area $l_0^2$ and it reads ${}^{R\!}\hat{E}_{(k)}\big(\mathbf{S}^{p}_v\big)\delta^k_p$.
Then the actions on the single cell state of the reduced gravitational operators
corresponding to \eqref{r_A} and \eqref{r_E} read respectively
\begin{equation}
\label{action_r_holonomy}
{}^{R}\hat{h}_{l^k_{\pm}}
\ket{c_v;j_l,i_v}_{\!R}
=
{}^{R}\hat{h}_{l^p}\delta^k_p
\ket{c_v;j_l,i_v}_{\!R}
={}^{R}h_{l^k_{\pm}}\!(v)
\ket{c_v;j_l,i_v}_{\!R}
=e^{\pm\varepsilon c_{(k)\!}(v)\tau^k\!/l_0}
\ket{c_v;j_l,i_v}_{\!R}
=:e^{\pm\mathbf{k}(v)}
\ket{c_v;j_l,i_v}_{\!R}
\end{equation}
and
\begin{equation}
\label{action_r_flux}
\hat{p}^{(k)}(\mathbf{S})
\ket{c_v;j_l,i_v}_{\!R}
=
\frac{1}{2}\Big({}^{R\!}\hat{E}_{(k)}\big(\mathbf{S}^{p}\big)+{}^{R\!}\hat{E}_{(k)}\big(\mathbf{S}^{-p}\big)\Big)\delta^k_p
\ket{c_v;j_l,i_v}_{\!R}
=\frac{\kbar}{2}\big(j^{(k)}_{v}+j^{(k)}_{v-\vec{e}_i}\big)
\ket{c_v;j_l,i_v}_{\!R}
=\kbar\Sigma^{(k)}_{v}
\ket{c_v;j_l,i_v}_{\!R}.
\end{equation}

In order to derive the quantity ${}^{R}\hat{O}\ket{c_v;j_l,i_v}_{\!R}$, we need to
calculate the action of the smeared reduced momentum on a holonomy, which reads
\begin{equation}
\hat{p}^i(v)h_{l^k}(v)
=-i\kbar\frac{\delta}{\delta c_{(i)}}e^{\mathbf{k}(v)}
=-i\kbar\frac{\delta}{\delta c_{i}}e^{\varepsilon c_{(k)\!}(v)\tau^{k}\!/l_0}
=-i\kbar\frac{\varepsilon}{l_0}\tau^ih_{l^k}(v),
\end{equation}
where $\hat{p}^i(v):=\hat{p}^{(i)}(\mathbf{S}_v)$ and
$h_{l^k}(v):=h_{l}(l^k_v)$.

Finally, we can derive the reduced Euclidean term,
$\epsilon^{ijk}\,\text{tr}\Big(\big(\hat{h}_{i\circlearrowleft
j}-\hat{h}_{i\circlearrowleft j}^{-1}\big)
\hat{h}_{l^k}^{-1}\,\hat{\mathbf{V}}\,\hat{h}_{l^k}\Big)$.
Considering for simplicity the fundamental representation of $\mathfrak{su}(2)$
with generators $\tau^i$, we have to evaluate the following action
\begin{equation}\label{Euclidean}
\epsilon^{ijk}\,\text{tr}
\Big[
\Big(
\hat{e^{\bf i}}\,\hat{e^{\bf j}}\,\hat{e^{-\bf i}}\,\hat{e^{-\bf j}}
-\hat{e^{\bf j}}\,\hat{e^{\bf i}}\,\hat{e^{-\bf j}}\,\hat{e^{-\bf i}}
\Big)
\hat{e^{-\bf k}}\,\hat{\mathbf{V}}\,\hat{e^{\bf k}}\Big]
\ket{c_v;j_l,i_v}_{\!R},
\end{equation}
where the loop hononomy has been decomposed into the line holonomies --- this, of course, can only be done since we have cuboidal graphs. The product under the trace is that of objects which are both geometrical operators acting on reduced states with an unchangeable order, as well as $\mathfrak{su}(2)$-valued elements obeying usual commutation relations and trace properties. Due to the fact that holonomy operator acts on $\ket{c_v;j_l,i_v}_{\!R}$ by multiplication, the relative order between the volume operator and the holonomies operators can be fixed only at the quantum level. In order to simplify the calculation, we first derive the formula
\begin{equation}\label{Ve}
{}^{R}\hat{\mathbf{V}}\,\hat{e^{\bf k}}
\ket{c_v;j_l,i_v}_{\!R}
=
{}^{R}\mathbf{V}_{\!v}\bigg(\mathds{1}+\frac{i\varepsilon}{2l_0\Sigma_v^{(k)}}\tau^k\Bigg)^{\!\!\frac{1}{2}}e^{\mathbf{k}(v)}
\ket{c_v;j_l,i_v}_{\!R}
=:
{}^{R}\textbf{\LARGE\textnu}_{\!v}^k\,e^{\mathbf{k}(v)}
\ket{c_v;j_l,i_v}_{\!R}.
\end{equation}
Hence the full Euclidean term gives
\begin{equation}\label{su2_loop1}
\begin{split}
&\epsilon^{ijk}\,\text{tr}\Big(\big(\hat{h}_{i\circlearrowleft
j}-\hat{h}_{i\circlearrowleft j}^{-1}\big)
\hat{h}_{l^k}^{-1}\,\hat{\mathbf{V}}\,\hat{h}_{l^k}\Big)
\ket{c_v;j_l,i_v}_{\!R}
\\=
&\ \epsilon^{ijk}\,\text{tr}\bigg(
\Big(
e^{\mathbf{i}(v)}\,e^{\mathbf{j}(v)}\,e^{-\mathbf{i}(v)}\,e^{-\mathbf{j}(v)}
-e^{\mathbf{j}(v)}\,e^{\mathbf{i}(v)}\,e^{-\mathbf{j}(v)}\,e^{-\mathbf{i}(v)}
\Big)
e^{-\mathbf{k}(v)}\,\textbf{\LARGE\textnu}_{\!v}^k\,e^{\mathbf{k}(v)}\bigg)
\ket{c_v;j_l,i_v}_{\!R}.
\end{split}
\end{equation}
This expression can be analytically solved using BCH formula. 
\\

Let us first focus on the loop part, $h_{i\circlearrowleft j}=e^{\bf i}\,e^{\bf
j}\,e^{-\bf i}\,e^{-\bf j}$. Using the following expansion:
\begin{equation}\label{BCH}
e^XYe^{-X}=e^{\,\text{ad}_X}Y=Y+[X,Y]+\frac{1}{2!}\big[X,[X,Y]\big]+\frac{1}{3!}\Big[X,\big[X,[X,Y]\big]\Big]+...,
\end{equation}
we can derive loops of both orientations,
\begin{equation}\label{BCH_e1}
{}^{R}h_{i\circlearrowleft j}
=
\mathds{1}+[{\bf i},{\bf j}]
+\frac{1}{2}\big[{\bf i},[{\bf i},{\bf j}]\big]
+\frac{1}{6}\Big[{\bf i},\big[{\bf i},[{\bf i},{\bf j}]\big]\Big]
+\mathcal{O}\big(\varepsilon^5\big),
\end{equation}
\begin{equation}\label{BCH_e2}
{}^{R}h_{i\circlearrowleft j}^{-1}
=
\mathds{1}-[{\bf i},{\bf j}]
+\frac{1}{2}\big[{\bf j},[{\bf i},{\bf j}]\big]
-\frac{1}{6}\Big[{\bf j},\big[{\bf j},[{\bf i},{\bf j}]\big]\Big]
+\mathcal{O}\big(\varepsilon^5\big).
\end{equation}
In the expansions above, the objects in the commutators are the exponents of the
exponential function in the definition of the reduced holonomy, precisely ${\bf
i}=\frac{\varepsilon}{l_0}c_{(i)}\tau^i$, ${\bf
j}=\frac{\varepsilon}{l_0}c_{(j)}\tau^j$. Then the commutators in \eqref{BCH_e1}
and \eqref{BCH_e2} read
\begin{equation}
\begin{split}
[{\bf i},{\bf j}]=&\
\epsilon_{ijk}\Big(\frac{\varepsilon}{l_0}\Big)^{\!2}c_{(i)}c_{(j)}\tau^k
,
\\
\big[{\bf i},[{\bf i},{\bf
j}]\big]=&\,-2\Big(\frac{\varepsilon}{l_0}\Big)^{\!3}c_{(i)}^2c_{(j)}\tau^j
,
\\
\big[{\bf j},[{\bf i},{\bf j}]\big]=&\
2\Big(\frac{\varepsilon}{l_0}\Big)^{\!3}c_{(i)}c_{(j)}^2\tau^i
,
\\
\Big[{\bf i},\big[{\bf i},[{\bf i},{\bf
j}]\big]\Big]=&\,-2\Big(\frac{\varepsilon}{l_0}\Big)^{\!4}c_{(i)}^2\epsilon_{ijk}c_{(i)}c_{(j)}\tau^k
,
\\
\Big[{\bf j},\big[{\bf j},[{\bf i},{\bf
j}]\big]\Big]=&\,-2\Big(\frac{\varepsilon}{l_0}\Big)^{\!4}c_{(j)}^2\epsilon_{ijk}c_{(i)}c_{(j)}\tau^k.
\end{split}
\end{equation}

The remaining part of \eqref{Euclidean} can be easily solved, using the fact that
commutator
$\big[{\bf k},\textbf{\LARGE\textnu}_{\!v}^k\big]$
is zero. Therefore we gets precisely
\begin{equation}\label{eVe}
e^{-\bf k}\,\textbf{\LARGE\textnu}_{\!v}^k\,e^{\bf
k}=\textbf{\LARGE\textnu}_{\!v}^k.
\end{equation}

Finally, imposing the $\mathfrak{su}(2)$ trace, we derive the full Euclidean
term, getting
\begin{equation}\label{trace_exp}
\begin{split}
\!\!\!\epsilon^{ijk}\,\text{tr}\Big(\big(\hat{h}_{i\circlearrowleft
j}-\hat{h}_{i\circlearrowleft j}^{-1}\big)
\hat{h}_{l^k}^{-1}\,\hat{\mathbf{V}}\,\hat{h}_{l^k}\!\Big)
\ket{c_v;j_l,i_v}_{\!R}
=&\ \epsilon^{ijk}\,\text{tr}\bigg[\bigg(
2\epsilon_{ijk}\Big(\frac{\varepsilon}{l_0}\Big)^{\!2}c_{(i)\!}(v)c_{(j)\!}(v)\tau^k
-\Big(\frac{\varepsilon}{l_0}\Big)^{\!3}c_{(i)\!}(v)c_{(j)\!}^2(v)\tau^i
\\
&-\Big(\frac{\varepsilon}{l_0}\Big)^{\!3}c_{(i)\!}^2(v)c_{(j)\!}(v)\tau^j
-\frac{1}{3}\Big(\frac{\varepsilon}{l_0}\Big)^{\!4}c_{(j)\!}^2(v)\epsilon_{ijk}c_{(i)\!}(v)c_{(j)\!}(v)\tau^k
\\
&-\frac{1}{3}\Big(\frac{\varepsilon}{l_0}\Big)^{\!4}c_{(j)\!}^2(v)\epsilon_{ijk}c_{(i)\!}(v)c_{(j)\!}(v)\tau^k
+\mathcal{O}(\varepsilon^5)\!
\bigg)\textbf{\LARGE\textnu}_{\!v}^k\bigg]
\ket{c_v;j_l,i_v}_{\!R}
\\
=&\,-\epsilon^{ijk}\epsilon_{ijk}c_{(i)\!}(v)c_{(j)\!}(v)
\frac{i\varepsilon^3}{4l_0^3\Sigma^{(k)}_{v}}\mathbf{V}_{\!v}
\\
&\times\bigg(\!1-\frac{1}{6}\Big(\frac{\varepsilon}{l_0}\Big)^{\!2}\big(c_{(i)\!}^2(v)+c_{(j)\!}^2(v)\big)
+\frac{\varepsilon^2}{2^7\big(l_0^2\Sigma^{(k)}_v\big)^{\!2}}+\mathcal{O}(\varepsilon^4)\!\bigg)
\ket{c_v;j_l,i_v}_{\!R},\!\!\!\!
\end{split}
\end{equation}
where we used the Taylor expansion around the value of the regulator $\varepsilon$. Considering the non-vanishing terms in expression \eqref{trace_exp}, up to the next to the leading order, we get the action of the gravitational Hamiltonian constraint operator,
\begin{equation}\label{q_r_A}
\begin{split}
\hat{H}^{(A)}
\ket{\Gamma;j_l,i_v}
=&\,
-\frac{2}{\gamma^2\kappa}\lim_{\varepsilon\to\varepsilon_0}\sum_{v\in\Gamma}\,l_0^3N(v)\sum_i^3
\frac{\mathbf{V}_{\!v}}{l_0^3\kbar\Sigma_v^{(i)}}\frac{c_1(v)c_2(v)c_3(v)}{c_{(i)\!}(v)}
\\&
\times\Bigg(1
-\Big(\frac{\varepsilon}{l_0}\Big)^{\!\!2}\,\frac{\,c_1^2(v)+c_2^2(v)+c_3^2(v)-c_{(i)\!}^2(v)}{6}
+\Big(\frac{\varepsilon}{l_0}\Big)^{\!\!2}\frac{1}{2^7\big(\Sigma_v^{(i)}\big)^{\!2}}
+\mathcal{O}(\varepsilon^4)
\Bigg)
\!\ket{\Gamma;j_l,i_v}.
\end{split}
\end{equation}
The latter precisely reproduces the results obtained by Alesci and Cianfrani \cite{Alesci:2014uha,Alesci:2015nja}.\\

It is worth mentioning that one can derive formula \eqref{trace_exp} exactly, \textit{i.e.} beyond a perturbative series. First, notice that multiplying formula $\epsilon^{ijk}\big(h_{i\circlearrowleft j}-h_{i\circlearrowleft j}^{-1}\big)$ with $\textbf{\LARGE\textnu}_{\!v}^k$, respectively, under the trace over the SU$(2)$ algebra, only odd terms of the first and even terms of the second formulas do not vanish. These terms can be gathered together, giving
\begin{equation}\label{trace_exact}
\begin{split}
\epsilon^{ijk}\,\text{tr}\Big(\big(h_{i\circlearrowleft j}-h_{i\circlearrowleft
j}^{-1}\big)\textbf{\LARGE\textnu}_{\!v}^k\Big)
=&
\,-\epsilon^{ijk}\epsilon_{ijk}\frac{i\varepsilon_0}{4l_0\Sigma^{(k)}_{v}}\mathbf{V}_{\!v}
\sum_{p,q,r=1}
\frac{(-1)^{p-1}}{(2p-1)!}\Big(\frac{\varepsilon_0}{l_0}c_{(i)\!}(v)\Big)^{\!2(p-1)}
\\
&\,\times
\frac{(-1)^{q-1}}{(2q-1)!}\Big(\frac{\varepsilon_0}{l_0}c_{(j)\!}(v)\Big)^{\!2(q-1)}
2^{5-4r}\bigg(\frac{\varepsilon_0}{l_0\Sigma_v^{(k)}}\!\bigg)^{\!\!2(r-1)}\binom{1/2}{2r-1}
\\
=&
\,-i\epsilon^{ijk}\epsilon_{ijk}\mathbf{V}_{\!v}
\sin\!\Big(\frac{\varepsilon_0}{l_0}c_{(i)\!}(v)\Big)\sin\!\Big(\frac{\varepsilon_0}{l_0}c_{(j)\!}(v)\Big)
\sum_{r=1}2^{5-4r}\bigg(\frac{\varepsilon_0}{l_0\Sigma_v^{(k)}}\!\bigg)^{\!\!2(r-1)}\binom{1/2}{2r-1}
\end{split}
\end{equation}
where $\sum_{p,q,r=1}$ runs over non-vanishing elements. Neglecting the inverse volume corrections, \textit{i.e.} taking the approximation
$\sum_{r=1}2^{5-4r}\big(l_0\Sigma_v^{(k)}/\varepsilon_0\big)^{\!2(1-r)}\binom{1/2}{2r-1}\approx1$,
we reproduce known result of LQC. Obviously, there is an important identification to be made at this point. In the standard notation of LQC, and as long as one sticks with the fundamental representation of SU($2$), the fraction $\varepsilon_0/l_0$ is proportional as $\mu_0$. In the work by Alesci and Cianfrani, it was shown that the $\bar{\mu}$ of LQC is related to the inverse number of nodes of the graph of the states, in which the expectation value of the Hamiltonian operator is taken in order to obtain the effective Hamiltonian. Indeed, in that case, if one works with a non-graph changing Hamiltonian, then this corresponds to the $\mu_0$ scheme rather than the `improved dynamics' of LQC. How does it measure up to our identification of $\bar{\mu}$ in our algebraic version of QRLG? The ratio of $\varepsilon_0/l_0$ tells us how big our cuboidal graphs are compared to the fundamental quantum scale. This ratio, in our case, is indeed dependent on the fiducial length and if it is made phase-space dependent, \textit{i.e.} $l_0$ is not treated as a constant any longer, then we can recover the $\bar\mu$ scheme as well. After all, that would imply having a non-fixed length of our fiducial cubes, which would correctly correspond to the graph-changing Hamiltonian of the Alesci-Cianfrani graphical picture.

\subsection{Matter sector}\label{IV.2}
\noindent
We solve the action of the trace of the reduced operator describing gravitational degrees of freedom coupled to matter, $\text{tr}\big(\tau^i\,{}^{R}\hat{h}_j^{-1}\,{}^{R}\hat{\mathbf{V}}^n\,{}^{R}\hat{h}_j\big)$, using a similar method to the one applied in section \eqref{IV.1}. Notice that the same expression has been used in the model of 2+1 gravity \cite{Bilski_2+1}. The BCH formula has been applied to the gravitational and the scalar field contribution to the Hamiltonian constraint. Let us repeat here the essential part of the calculations performed in \cite{Bilski_2+1}. Using the expressions \eqref{Ve} and \eqref{eVe}, we get
\begin{equation}\label{su2_line2}
\begin{split}
\,\text{tr}\Big(\tau^i\hat{h}_{l^j}^{-1}\,\hat{\mathbf{V}}^n\,\hat{h}_{l^j}\Big)
\ket{\Gamma;j_l,i_v}
=&\ \,\text{tr}\big(
\tau^i\,(\textbf{\LARGE\textnu}^j_{\!v})^n\big)
\ket{\Gamma;j_l,i_v}
\\
=&\ -\frac{i\varepsilon
n\,\delta^i_j}{2^3l_0\Sigma^{(j)}_v}\big(\mathbf{V}_{\!v}\big)^{\!n}
\Bigg(1+\varepsilon^2\,\frac{(n-2)(n-4)}
{2^7\!\cdot\!3\big(l_0\Sigma^{(j)}_v\big)^{\!2}}+\mathcal{O}(\varepsilon^4)\Bigg)\!
\ket{\Gamma;j_l,i_v}.
\end{split}
\end{equation}
We can rewrite the latter into the series form
\begin{equation}\label{su2_matter}
\begin{split}
\,\text{tr}\big(
\tau^i\,(\textbf{\LARGE\textnu}^j_{\!v})^n\big)
=
-\frac{i\varepsilon
n\,\delta^i_j}{2^3l_0\Sigma^{(j)}_v}\big(\mathbf{V}_{\!v}\big)^{\!n}
\sum_{r=1}\frac{2^{5-4r}}{n}\bigg(\frac{\varepsilon}{l_0\Sigma_v^{(j)}}\!\bigg)^{\!\!2(r-1)}\binom{n/2}{2r-1}.
\end{split}
\end{equation}

Applying this formula respectively to the scalar and vector fields, smeared in
the lattice representations proposed by Thiemann \cite{Thiemann:1997rt,Thiemann:2007zz}, we get respectively
\begin{equation}\label{q_r_phi}
\begin{split}
\!\!\!\hat{H}^{(\phi)}\!
\ket{\Gamma;j_l,i_v;\text{matter}}
=&\ \frac{1}{2\zeta}\lim_{\varepsilon\to\varepsilon_0}\sum_{v\in\Gamma}N(v)
\Bigg\{
\zeta^2\pi^A(v)\pi^A(v)
\frac{\big(\mathbf{V}_{\!v}\big)^{\!3}}{\kbar^6\big(\Sigma^1_v\Sigma^2_v\Sigma^3_v\big)^{\!2}}
\!\Bigg[1+\Big(\frac{\varepsilon}{l_0}\Big)^{\!\!2}\,\frac{21}{2^8}
\Bigg(\frac{1}{\big(\Sigma^1_v\big)^{\!2}}+\frac{1}{\big(\Sigma^2_v\big)^{\!2}}+\frac{1}{\big(\Sigma^3_v\big)^{\!2}}\Bigg)
\Bigg]\!\!\!\!
\\
&+
\sum_i^3
\frac{l_0^2\big(\Sigma^{(i)}_v\big)^{\!2}\big(\mathbf{V}_{\!v}\big)^{\!3}}{\kbar^4\big(\Sigma^1_v\Sigma^2_v\Sigma^3_v\big)^{\!2}}
\,\partial_{(i)}\phi^A(v)\,\partial_{(i)}\phi^A(v)
\!\Bigg[1+\Big(\frac{\varepsilon}{l_0}\Big)^{\!\!2}\,\frac{65}{2^{9}\!\cdot\!3}
\Bigg(\frac{1}{\big(\Sigma^1_v\big)^{\!2}}+\frac{1}{\big(\Sigma^2_v\big)^{\!2}}+\frac{1}{\big(\Sigma^3_v\big)^{\!2}}\Bigg)
\Bigg]
\\
&+
\mu^2\,\mathbf{V}_{\!v}\,\phi^A(v)\,\phi^A(v)
\Bigg\}
\ket{\Gamma;j_l,i_v;\text{matter}}
\end{split}
\end{equation}
and
\begin{equation}\label{q_r_underline}
\begin{split}
\hat{H}^{(\underline{A})}
\ket{\Gamma;j_l,i_v;\text{matter}}
=&\
\frac{l_0^4}{2^3Q^2}\lim_{\varepsilon\to\varepsilon_0}\sum_{v\in\Gamma}N(v)\sum_i^3
\frac{\mathbf{V}_{\!v}}{\kbar^2\big(\Sigma^{(i)}_v\big)^{\!2}}
\Bigg\{
\,Q^4\underline{E}^{(i)}_A(v)\,\underline{E}^{(i)}_A(v)
+
\underline{B}^{(i)}_A(v)\,\underline{B}^{(i)}_A(v)
\Bigg\}
\\
&\times
\Bigg[1+\Big(\frac{\varepsilon}{l_0}\Big)^{\!\!2}\frac{7}{2^8}
\Bigg(\frac{1}{\big(\Sigma^1_v\big)^{\!2}}+\frac{1}{\big(\Sigma^2_v\big)^{\!2}}+\frac{1}{\big(\Sigma^3_v\big)^{\!2}}\Bigg)
\Bigg]
\ket{\Gamma;j_l,i_v;\text{matter}}.
\end{split}
\end{equation}
Notice that we did not defined here the lattice representation for the matter field variables. We only considered a simple lattice node expansion $f=\sum_vf(v)$ of field variable $f$. Detailed derivations of the matter field actions and definitions of the lattice representations for the matter field operators can be found in \cite{Bilski:2015dra,Bilski:2016pib,Bilski_Thesis}. In this case, and differently than in the previous section where we also calculated the non-perturbative expressions, we restricted our expansions to the leading terms, which reproduce the classical expressions \eqref{T_scalar} and to the next to the leading order corrections. It is worth mentioning that these expansions have been performed around a small value of regulator $\varepsilon$, after the introduction of fiducial length $l_0$ in \eqref{r_A} and \eqref{r_E}, which is independent of regulator $\varepsilon$ (however its value is bounded from below, $l_0\ge\varepsilon_0$ --- see \ref{IV.2}). Then, the semiclassical limit can be reproduced without any assumptions on the value of spin $j$ (in particular, without using the large-$j$ expansion). This shall be made more explicit in forthcoming work \cite{Bilski_et_al.}. Finally, we omit calculations concerning the fermionic sector of matter, which are in preparation.

\section{Summary and Discussions}\label{V}

\noindent
In this section, we conclude by investigating why the new method works in QRLG. We first begin by a brief summary of our results. We have rederived the effective Hamiltonian operator in QRLG using customary tricks deployed in LQC. This way we reinforce the close relationship between the two approaches while simultaneously uncovering a much simpler short-cut of deriving the effective Hamiltonian in models of QRLG. This also helps in coupling matter fields in this formulation, as dealing with the volume operator can be shown to be considerably more manageable in this case. However, we also raise important issues while going through our algorithm. Firstly, how can we recover the continuum limit starting from such a granular spacetime, as the one introduced in QRLG? Although such discretization is rather helpful for several technical reasons, one has to ensure that GR is recovered in the continuum case. This is a question that we would like to answer in the future with a particular focus on why the regulator cannot simply be taken to zero but rather to a physical constant, related to the Planck length.

Another important question which comes up along the way, concerns the relationship between LQC and QRLG. Although this has been investigated in some detail in the past \cite{Alesci:2015nja,Alesci:2016gub}, we approach this issue from a different perspective here. After defining the basic canonical variables along the edges and vertices of our fundamental cuboids, we show how the methods of LQC can be applied to the case of RLQG in order to derive the regularized Hamiltonian operator and so, in effect, its dynamics. Superficially, it is not easy to discern why the recoupling theory in terms of reduced cuboidal graphs would be the same as that of employing the algebraic methods of LQC. However, on closer inspection we notice the deeper connection between the two, which is elaborated upon in the next subsection.

\subsection{Abelian connections: why the shortcut works}\label{IV.2}
\noindent It is interesting to investigate further why our method reproduces the results obtained in QRLG using cuboidal graphs. \`A priori, there does not seem to be an obvious reason for that. On the one hand, one calculates matrix elements in QRLG using complicated graphs after using some gauge fixings. On the other hand, in our method, one replaces point-wise holonomies in the operators before applying them on coherent states, without directly resorting to the graph structure of the underlying spin-networks. The reason for this lies precisely in the gauge-fixings used in QRLG. The choice of a \textit{diagonal} tetrad field, in QRLG, mandates that the internal  SU$(2)$ group of rotations breaks into the product of three U$(1)$ groups. Each of the SU$(2)$ elements is projected into the three orthogonal directions, each labeled by an abelian U$(1)$ group\footnote{Since we are dealing with an isotropic reduction in this case, unlike that of Bianchi models usually done in QRLG, our reduced gauge group is $U(1)^3$.}. This is the crucial simplification used in the theory. It is worth emphasizing once again, that one does not choose such an abelian group \textit{ad hoc}, but is rather dictated to do so due to the symmetric gauge fixing imposed on the physical tetrad field.

Before we explore this in detail, it is important to remember that an Abelian connection is not the same as a homogeneous one. Abelian connections have a rather unique feature absent in their non-abelian counterparts as has been pointed out in detail in \cite{BojowaldLQCMath}. Quantizations of Abelian connections are based on functions defined on the Bohr compactification of the real line. For instance, U$(1)$-holonomies based on an isotropic connection, $c$, maybe written as
\begin{eqnarray}
  h^{\bar{\mu}}=\exp\left(i \bar{\mu} c\right),
\end{eqnarray}
with $\bar{\mu} \in \mathbb{R}$.
This $\bar{\mu}$ maybe thought of an artifact coming from the full theory in the form of an area-gap parameter. The holonomies are usually calculated along all pieces along the edges of the integration cube of lengths $\varepsilon_0 \le l_0$, with $l_0^3 := \int \text{d}^3x$ being the coordinate volume of the region under consideration.
However, if one considers integer representations of the abelian U$(1)$ group, $n$, then this parameter $\bar{\mu}$\footnote{Whether we work in the constant $\mu$ or the improved-$\bar{\mu}$ scheme is irrelevant for this purpose.} may be thought of as the product of the representation label $n$ and the fractional edge length $\lambda = \varepsilon_0/l_0$. Thus, one can span the Hilbert space of all integrable functions of the Bohr-compactified real line, $\bar{\mathbb{R}}_\text{Bohr}$, rather than than some periodic interval of $\mathbb{R}$. This is what is usually used extensively in constructions of LQC, and is at the heart of our point-wise holonomy operators used in our analysis.

Nonetheless, there is an identity which is used implicitly in this construction, which is only true for abelian systems. A U$(1)$-holonomy $\exp(iλc)$, calculated in the $n$-representation, may be written as
\begin{eqnarray}
  \rho_n\left(\exp(i\lambda c)\right) = \exp(i\lambda n c)\,.
\end{eqnarray}
On the other hand, as a representation of $\bar{\mathbb{R}}_\text{Bohr}$, one may
evaluate the $\lambda n$-representation as
\begin{eqnarray}
  \rho_{\lambda n}\left(\exp(ic)\right) = \exp(i \lambda nc).
\end{eqnarray}
These two functions of $c$ agrees and can be identified with each other. This agreement of the functional values is usually utilized in LQC to construct the states on Hilbert space, as functions on $\bar{\mathbb{R}}_\text{Bohr}$. With a simple counterexample, we can show that this fortuitous mathematical property only exists for Abelian connections. For a SU$(2)$ connection the two functions $\rho_j \exp(i\lambda c)$ and $\rho_{\lambda j} \exp(ic)$ not only are not equal but the latter expression is not even defined when $\lambda$ is an integer or half-integer. In this analysis, it is important to remember that one merges two different quantities to obtain the parameter $\bar{\mu}$, the edge length of the spin network state and the representation label.

This is what is also done for QRLG. One first defines a cuboidal graph, and then gauge-fixes the tetrad to break the SU$(2)$ into three U$(1)$-groups, the crucial point being the fact that the residual symmetry groups are Abelian. Elements of the reduced kinematical Hilbert space are labelled by cuboidal graphs, having U$(1)$ group elements at links and some reduced intertwiners at nodes, which are labelled by complex numbers. In practise, this is then the same as what is done for LQC and that is why our method matches with that of QRLG.

\vspace{1cm}

{\it{\textbf{Acknowledgements}}}\\
\noindent
J.B. and A.M. wishes to acknowledge support by the Shanghai Municipality, through the grant No. KBH1512299, and by Fudan University, through the grant No. JJH1512105.



\end{document}